\documentclass[5p,twocolumn,times,number]{elsarticle}

\usepackage{graphicx}
\usepackage{amsmath}   
\def\pmbanner{{\hrule height 1 pt}\vskip35pt{NIMA POST-PROCESS BANNER TO BE REMOVED AFTER FINAL ACCEPTANCE}\vskip35pt{\hrule height 4pt}\vskip20pt}

\begin{document}

\begin{frontmatter}

  \title{\pmbanner Implementation of the code for the simulation of the response of a triple-GEM tracker and its comparison to the experimental data}

  \author[a]{L.~Lavezzi\corref{cor}} \ead{lia.lavezzi@to.infn.it}
  \author{on behalf of the CGEM-IT group} 

\cortext[cor]{Corresponding author}

\address[a]{Institute of High Energy Physics, Beijing, China {\it and} Istituto Nazionale di Fisica Nucleare, Sezione di Torino, Italy}


\begin{abstract}
In the framework of detector development, Monte Carlo simulations play a key role in the evaluation of the expected performance and the full understanding of the behavior in beam conditions. In particular, a software which simulates the response of the detector to the particle passage is mandatory to test different setups and solutions, such as geometries, fields, voltages etc. and to understand the test beam data. For gas trackers, existing softwares, such as GARFIELD, perform a very detailed simulation of the physical processes but are also CPU time consuming. For the new cylindrical GEM tracker of BESIII, a faster code which models the results obtained from GARFIELD and adapts them to the experimental data, collected in several test beams, has been written. It reproduces the behavior of a planar triple-GEM under different working conditions and, when completed, it will be inserted in the official code of BESIII. A description of the procedure, based on different components (ionization, diffusion and magnetic field, avalanche multiplication, signal induction and readout) will be given and its results will be compared to the GARFIELD simulations and to the experimental data.
\end{abstract}


\end{frontmatter}

\section{The Gas Electron Multiplier digitization}
The {\it digitization} is the simulation of the response of the detector to the passage of the particle of interest. For Gas Electron Multipliers (GEM, \cite{gem}), the most used tool is GARFIELD \cite{garfield}. It is, however, CPU time consuming and cannot be applied in the global framework of an experiment as BESIII \cite{bes3}. For the CGEM-IT \cite{cgemit} we developed a standalone code which provides a full treatment of the detector response, from the passage of the particle to the signal formation, in a fast and reliable way. Once completed, it will also be ported to BESIII framework. \\
The starting point of the implementation was the article by W. Bonivento {\it et al.} \cite{bonivento}. Since different characteristics of the fields, the gas and the geometry are independent, they can be evaluated separately and the pieces of code describing each of them can be merged together on a second stage. Four steps were foreseen: ionization, GEM properties description, gas and magnetic field property effects, signal formation and noise. In each step, the parameters were evaluated by dedicated GARFIELD simulations and were used to model the development of the electron cloud from the drift gap up to the anode.
\subsection{Ionization}
The number of the primary ionization clusters follows Poisson statistics, thus the distance between two consecutive ionization points can be sampled from an exponential distribution. Both the mean number of primary ionizations and of secondary electrons in each cluster were evaluated directly from GARFIELD simulations.
\subsection{GEM properties}
A GEM can be described in terms of its {\it gain}, i.e. the multiplication factor of each generated electron after passing through the GEM holes, and of its {\it transparency}, i.e. the collection efficiency multiplied by the extraction efficiency. The former depends on the applied high voltage, the latter on the fields between two electrodes. These were evaluated in GARFIELD at different high voltage and field settings and the behavior of each GEM was simulated accordingly in the standalone digitization code. To obtain the parameters, a set of electrons was generated on a plane placed $150$ $\mu$m before each GEM and drifted through the holes to a plane $150$ $\mu$m after the GEM, with the gain simulation switched on.
\subsection{Effect of the gas}
Transverse and longitudinal diffusion effects due to the multiple scattering of electrons on the gas molecules have been evaluated with GARFIELD. Their effect is an enlargement of the charge distribution on the anode, which was evaluated by drifting a set of electrons in each of the gas gaps, where different electric fields are set, from one electrode to the next one and measuring the entity of the enlargement of the electronic cloud on the final plane. In the standalone simulation the position of each electron is sampled from a Gaussian distribution with mean equal to the starting position and sigma defined by the combination of the contrbutions of the different gaps. There is also the possibility to switch on the magnetic field: it shifts the mean value of this distribution according to the Lorentz force deviation. The longitudinal diffusion provides a spread and a shift in the distribution of the time of arrival of the different signals on the strips. This was evaluated in GARFIELD and implemented in the standalone code in a similar way.
\subsection{Signal formation}
The instantaneous current induced on a strip by an electron can be computed by the Shockley-Ramo theorem \cite{ramo} as
\begin{equation}
i_{ind} = q_e \cdot \overline{v}_{\textrm{drift}} \cdot \overline{W}
\end{equation}
where $q_e$ is the electron charge, $\overline{v}_{\textrm{drift}}$ is the drift velocity and $\overline{W}$ is the so-called {\it weighting field}. The weighting field is the electric field generated by the electrode under consideration when kept at $1$ V with all the other electrodes set to $0$ V. It was computed from the electric potential calculated by means of the finite element method.
\section{Tuning of the simulation to the experimental data}
\begin{figure}[!t]
\centering
\includegraphics[width=0.99\linewidth]{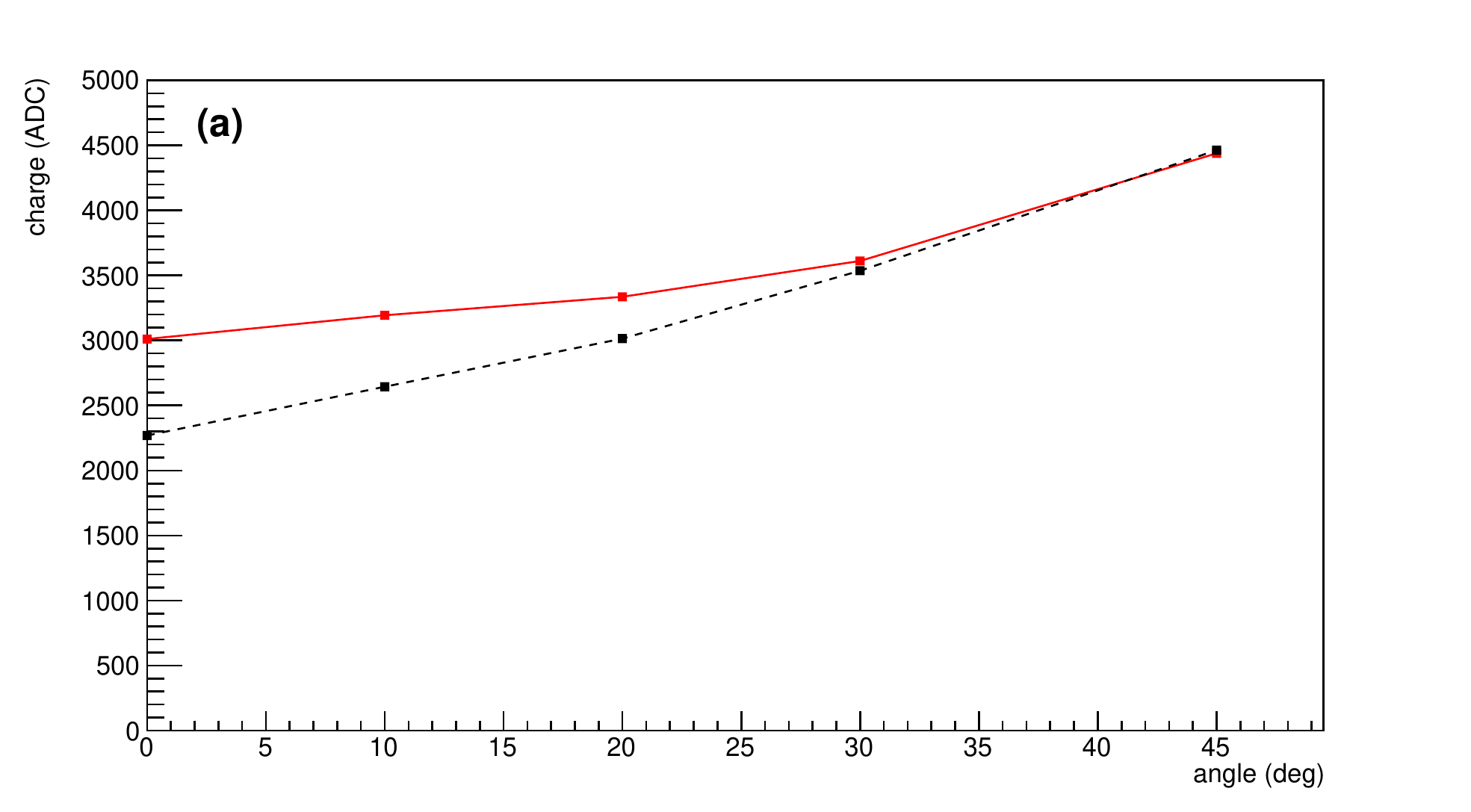}
\includegraphics[width=0.99\linewidth]{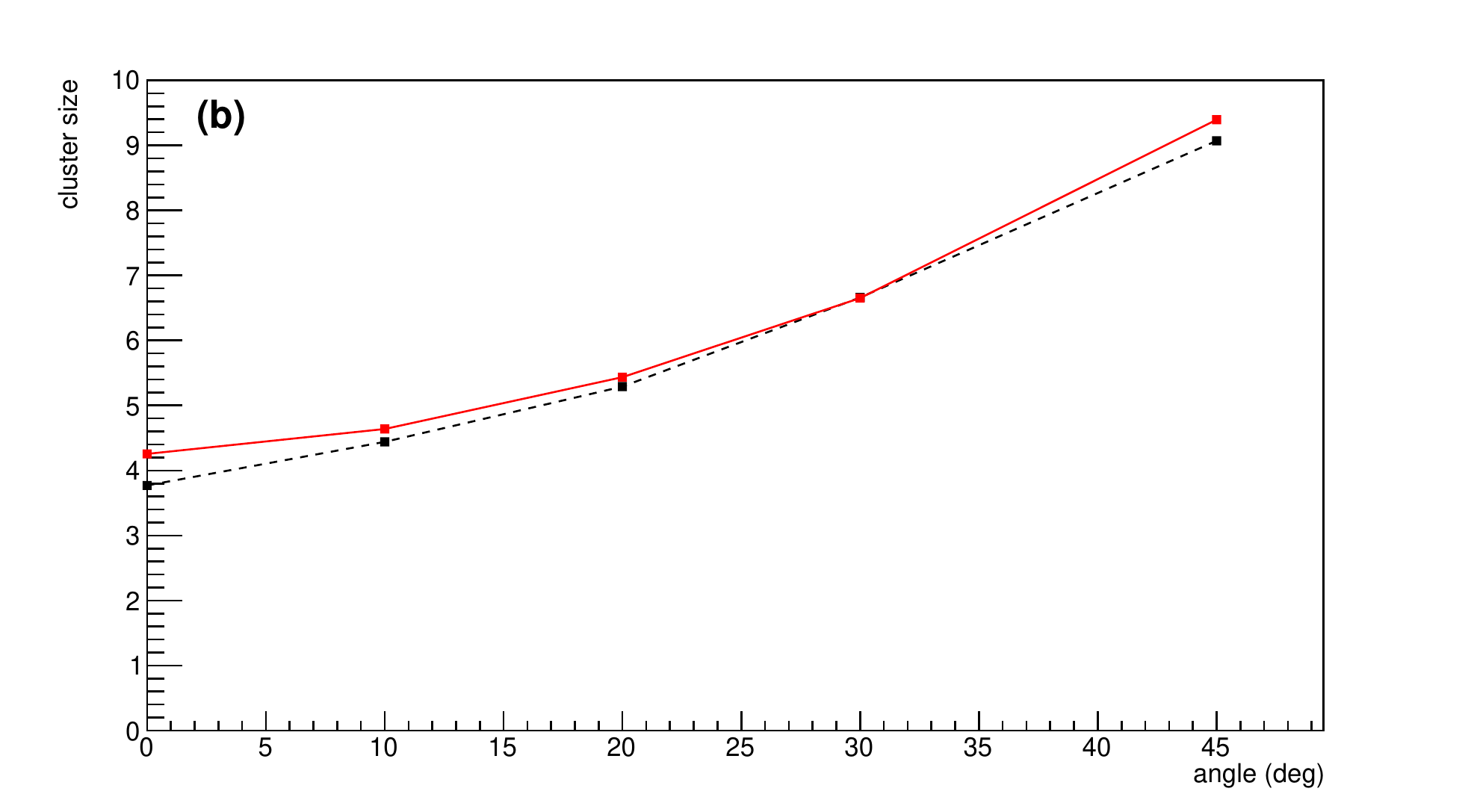}
\includegraphics[width=0.99\linewidth]{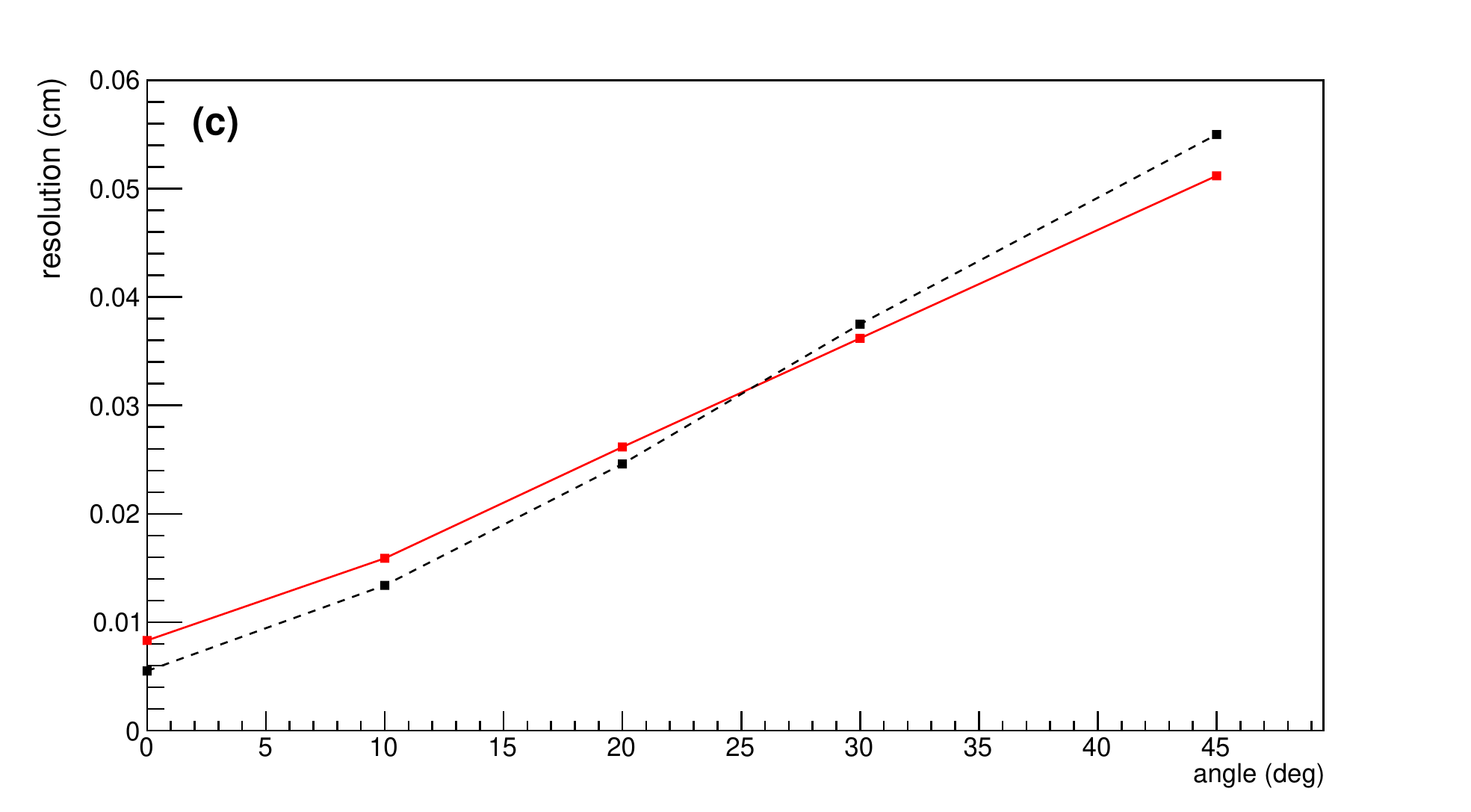}
\caption{Comparison of simulation (red full line) and test beam (black dotted line). (a) charge (ADC counts), (b) cluster size, (c) charge centroid position resolution (cm) {\it vs} incident angle (degree), without magnetic field.} \label{fig:tuning}
\end{figure}

Eventually, the aim of the digitization is to obtain simulated data which resemble the experimental ones (within a certain tolerance) in order to foresee the performance which will be delivered by the detector under development once finished. In order to obtain reliable results, the digitization first requirement is to reproduce the test beam results. A scan of different values of the angle of incidence of the beam was performed and the readout charge, the cluster size and the resolution obtained with the charge centroid reconstruction were evaluated \cite{tb}. In figure \ref{fig:tuning} the obtained experimental values of the three variables are compared to the simulated ones inside the standalone code. For the charge, an agreement around $20 \%$ was found, for the cluster size around $15 \%$ and for the position resolution around $20 \%$. The agreement is very good but can be improved, moreover since the final reconstruction will foresee also the use of the micro-TPC method \cite{mutpc}, the time resolution and the $\mu-$TPC position resolutions will be tuned to the experimental data. The agreement will be evaluated also in magnetic field.
\section*{Acknowledgments}
The research leading to these results has been performed within the BESIIICGEM Project, funded by the European Commission in the call H2020-MSCA-RISE-2014.

\end{document}